\documentclass[journal]{IEEEtran}
\usepackage[OT1]{fontenc}
\usepackage[numbers, sort&compress]{natbib}
\usepackage[cmex10]{amsmath}
\usepackage{amssymb}
\usepackage{bm}
\usepackage{braket}
\usepackage{booktabs}
\usepackage{graphicx, xcolor}
\usepackage{color}
\usepackage{subfigure}
\usepackage{tabularx}
\usepackage{arydshln}
\usepackage{mathtools}
\usepackage{multirow}
\usepackage{algorithm, algorithmic}
\usepackage{url}
\usepackage{listings}
\usepackage{CJKutf8}
\usepackage{comment}
\usepackage{physics}
\usepackage{hyperref}
\usepackage[absolute,overlay]{textpos}

\def\CN{\mathcal{CN}}

\def\s{\mathbf{s}}

\def\c{\mathbf{c}}

\def\E{\mathbb{E}}
\def\p{\mathbf{p}}

\def\y{\mathbf{y}}

\def\C{\mathbb{C}}
\def\n{\mathbf{n}}

\newcommand{\minimize}{\mathop{\rm minimize}\limits}

\RequirePackage[normalem]{ulem} 
\RequirePackage{color}\definecolor{RED}{rgb}{1,0,0}\definecolor{BLUE}{rgb}{0,0,1}

\makeatletter
\def\ps@IEEEtitlepagestyle{%
  \def\@oddfoot{\mycopyrightnotice}%
  \def\@oddhead{\hbox{}\@IEEEheaderstyle\leftmark\hfil\thepage}\relax
  \def\@evenhead{\@IEEEheaderstyle\thepage\hfil\leftmark\hbox{}}\relax
  \def\@evenfoot{}%
}
\def\mycopyrightnotice{%
  \begin{minipage}{\textwidth}
  \centering \scriptsize
\textcopyright 2025 IEEE. Personal use of this material is permitted.
  Permission from IEEE must be obtained for all other uses, in any current or future
  media, including reprinting/republishing this material for advertising or promotional
  purposes, creating new collective works, for resale or redistribution to servers or
  lists, or reuse of any copyrighted component of this work in other works.
  DOI: \href{https://doi.org/10.1109/LWC.2025.3540043}{10.1109/LWC.2025.3540043}
  \end{minipage}
}
\makeatother
\begin{document}
\title{\LARGE Covert Communications Without Pre-Sharing of Side Information and Channel Estimation Over Quasi-Static Fading Channels}

\author{
    Hiroki~Fukada,~\IEEEmembership{Graduate Student Member,~IEEE},
    Hiroki~Iimori,~\IEEEmembership{Member,~IEEE},\\
    Chandan~Pradhan,~\IEEEmembership{Member,~IEEE},
    Szabolcs~Malomsoky,
    and Naoki~Ishikawa,~\IEEEmembership{Senior Member,~IEEE}.
    \thanks{H.~Fukada and N.~Ishikawa are with the Faculty of Engineering, Yokohama National University, 240-8501 Kanagawa, Japan (e-mail: ishikawa-naoki-fr@ynu.ac.jp). H.~Iimori, C.~Pradhan, and S.~Malomsoky are with Ericsson Research, Ericsson Japan K.K., 220-0012 Kanagawa, Japan.
    }}

\markboth{} {Shell \MakeLowercase{\textit{et al.}}: Bare Demo of IEEEtran.cls for Journals}
\maketitle
\TPshowboxesfalse
\begin{textblock*}{\textwidth}(45pt,10pt)
\footnotesize
\centering
Accepted for publication in IEEE Wireless Communications Letters. This is the author's version which has not been fully edited and content may change prior to final publication. Citation information: DOI 10.1109/LWC.2025.3540043
\end{textblock*}

\begin{abstract}
    We propose a new covert communication scheme that operates without pre-sharing side information and channel estimation, utilizing a Gaussian-distributed Grassmann constellation for noncoherent detection. By designing constant-amplitude symbols on the Grassmann manifold and multiplying them by random variables, we generate signals that follow an arbitrary probability distribution, such as Gaussian or skew-normal distributions. The mathematical property of the manifold enables the transmitter's random variables to remain unshared with the receiver, and the elimination of pilot symbols that could compromise covertness. The proposed scheme achieved higher covertness and achievable rates compared to conventional coherent Gaussian signaling schemes, without any penalty in terms of complexity.
\end{abstract}

\begin{IEEEkeywords}
    Covert communications, Grassmann manifold, noncoherent detection, quasi-static
    fading channels.
\end{IEEEkeywords}

\IEEEpeerreviewmaketitle

\section{Introduction}

\IEEEPARstart{C}{overt} communication aims to convey confidential information
without revealing the existence of the communication itself to third parties~\cite{chen2023covert}.
This concept has been investigated for a multitude of communication systems,
including military, cellular networks, IoT, unmanned aerial vehicles~(UAVs), and satellite networks~\cite{Jiang2024covertsurvey}.
The performance of covert communications has been studied from theoretical as well as practical perspectives.
It has been shown that when the input to an additive white Gaussian noise (AWGN)
channel follows a Gaussian distribution,
the information rate is proportional
to the square root of the Gaussian channel length \cite{bash2013limitsSISO},
and the covertness is maximized \cite{yan2019gaussian}.
In parallel, the chaos-based modulation scheme has been proposed as an efficient approach for generating Gaussian-distributed signals~\cite{2012OkamotoChaos},
which can be considered suitable for covert communications.

For practical covert communication scenarios,
channel estimation remains indispensable but poses a critical problem,
as the frequent transmission of the pilots is required to improve the estimation accuracy, 
but simultaneously, 
it can be exploited by the third parties  to estimate their channel, 
thereby leading to a significant reduction in covertness~\cite{yan2019}.
For such a problem,
Katsuki et al. have proposed a new noncoherent communication system based on differential encoding~\cite{katsuki2023NGS},
where the projection onto the Gaussian distribution generates Gaussian signaling:
however, for decoding,
it requires at least one substantial reference symbol block.
In addition, 
similar to the chaos-based modulation~\cite{2012OkamotoChaos}, 
it necessitates the perfect sharing of  
a side information related to the projection in advance, which has been an open issue in the design of practical systems.

To address the limitations of conventional studies,
we propose a noncoherent
covert communication scheme,
where the information can be decoded without channel estimation as well as side information~(e.g., 
the chaos transition~\cite{2012OkamotoChaos} and the projection matrix~\cite{katsuki2023NGS}) that generates Gaussian signals at legitimate transceivers.
Unlike other benchmarks, 
our approach eliminates the necessity of
transmitting any pilot symbols and sharing side information,
making covert communications more practical. 
The major contributions are twofold.
\begin{enumerate}
    \item \textit{We design Grassmann symbols distributed on a unit circle and generate symbols that follow a Gaussian distribution} by multiplying them with Gaussian-distributed random variables. This process does not hinder noncoherent detection and requires no sharing of side information. Additionally, it can support symbol generations following an arbitrary probability distribution.\footnote{In terms of mutual information, the Gaussian signaling can be outperformed by the skew-normal signaling \cite{yan2019gaussian}, which is supported by our scheme.}
    \item \textit{We compare the proposed noncoherent scheme with two coherent detection-based schemes considered as benchmarks, demonstrating a higher covertness.} Then, we analyze the channel estimation errors that occur when transmitting Gaussian-distributed pilot symbols, showing that the proposed scheme also excels in terms of achievable rate.
\end{enumerate}

\section{System Model}
\label{sec:sys}
A quasi-static Rayleigh fading channel is considered in this letter.
A legitimate transmitter, Alice, communicates with a legitimate receiver, Bob, and a warden, Willie, tries to detect the communication.
We assume that the channel matrices between Alice-Bob and Alice-Willie are perfectly uncorrelated.

\subsection{Alice-Bob Channel}

Firstly, we consider coherent detection systems.
Let $\mathcal{H}_{1}$ denote the event that Alice transmits information,
and $\mathcal{H}_{0}$ the event that she does not.
For simplicity, we assume the perfect timing synchronization between the legitimate transceiver.

\begin{figure}[tb]
    \centering
    \includegraphics[width=1.00\linewidth, keepaspectratio,clip]
    {
        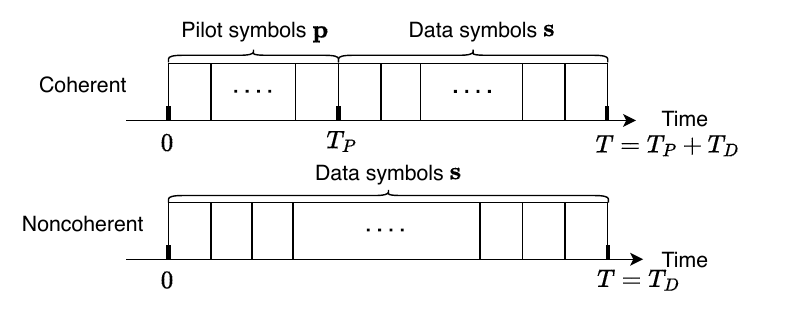
    }
    \vspace{-1ex}
    \caption{Frame structure assumed in coherent and noncoherent systems.}
    \vspace{-2ex}
    \label{fig:frame}
\end{figure}
We focus on the event with $\mathcal{H}_{1}$.
As illustrated in Fig.~\ref{fig:frame}, a frame consists of $T_{P}$ symbols $\p = \qty[p_{1} ~ \cdots ~ p_{T_{P}}] \in \C^{T_P}$ for pilot transmission and $T_{D}$ symbols $\s = \qty[s_{1} ~ \cdots ~ s_{T_{D}}] \in \C^{T_{D}}$ for data transmission carrying $B$ information bits.
Let $T \triangleq T_{P} + T_{D}$, and the transmit symbol vector is expressed as $[\p ~ \s] \in \C^{T}$.
The received symbol vector at Bob is given by
\begin{align}
    \y_{b}= \s h_{b}+\n_{b}, \label{eq:yb}
\end{align}
where $h_{b} \sim \mathcal{CN}(0, 1)$ denotes the channel coefficient of Bob, and $\n_{b}$ denotes an AWGN vector with each element following $\CN\qty(0,\sigma^{2}_{b})$.
The received signal-to-noise ratio~(SNR) at Bob is then defined as
$\gamma_{b}\triangleq \E[\norm{\s h_ b}^{2}]/\E[\norm{\n_{b}}^{2}]=1/\sigma^{2}_{b}$.

Bob estimates his own channel by the minimum mean square error~(MMSE) equalizer.
Let $\y_{b,P} \in \C^{T_{P}}$ denote the received symbols corresponding to the pilot symbols, and the estimated channel is then expressed as
$\hat{h}_{b}=\y_{b,P} (\norm{\p}^{2}+\sigma^{2}_{b})^{-1}\p^{\mathrm{H}}$,
with $\qty(\cdot)^{\mathrm{H}}$ representing the conjugate transpose.
Note that the pilot symbols used for channel estimation
have to be shared with the legitimate receiver before data transmission.

\subsection{Detection Error Probability at Willie}
\label{subsec:dep}
Willie monitors the Alice-Bob communication channel
and aims to detect the existence of their communication.
We assume that Willie can perform perfect synchronization
such that the received signal is sampled at the Nyquist rate.
The received symbol vector is $\y_{w}=\n_{w}$ for $\mathcal{H}_{0}$ and $\s h_{w}+\n_{w}$ for $\mathcal{H}_{1}$,
where $h_{w} \sim \mathcal{CN}(0, 1)$ denotes the channel coefficient of Willie and $\n_{w}$ denotes
an AWGN vector.
The received SNR at Willie is defined as
$\gamma_{w}\triangleq \E [\norm{\s h_{w}}^{2}]/\E[\norm{\n_{w}}^{2}]=
    1/\sigma^{2}_{w}$.

Assuming that $\s$ consists of independent and identically distributed~(i.i.d) Gaussian random variables~(RVs) of length $T$,
Willie estimates the existence of communication with a threshold $\lambda$ and the radio meter detection~\cite{Shah2021fading}
\begin{align}
    \frac{\norm{\y_{w}}^{2}}{T}\overset{\hat{\mathcal{H}}_{1}}{\underset{\hat{\mathcal{H}}_{0}}{\gtrless}}\lambda ,\label{ineq:Testing}
\end{align}
where $\hat{\mathcal{H}}_{i}$ with $i\in\qty{0,1}$ is the estimated event by Willie.\footnote{An exhaustive search-based strategy can be considered ideal, but we expect it decreases DEPs for coherent and noncoherent schemes almost equally.}

Under the assumption that both events $\mathcal{H}_{0}$ and $\mathcal{H}_{1}$ occur with an equal
probability,
the detection error probability (DEP) at Willie is given by
$P_{e,w} (\lambda) = (P_{\text{FA}} (\lambda) + P_{\text{MD}} (\lambda)) / 2$~\cite{bash2013limitsSISO}, where $P_{\text{FA}}\qty(\lambda)$ and $P_{\text{MD}}\qty(\lambda)$ are the probabilities
of the false alarm and the miss detection, respectively.
Both are expressed as $P_{\text{FA}}\qty(\lambda) \triangleq P (\hat{\mathcal{H}}_{1}\mid \mathcal{H}_{0}) = 1 - \gamma (T, T\lambda / \gamma^{-1}_{w}) / (T - 1)!$ and $P_{\text{MD}} (\lambda) \triangleq P (\hat{\mathcal{H}}_{0}\mid \mathcal{H}_{1}) = \gamma (T, T\lambda / (\abs{h_{w}}^{2}+ \gamma^{-1}_{w})) / (T - 1)!.$
Here, the lower incomplete Gamma function is given by $\gamma(x, a) = \int^{a}_{0}t^{x-1}e^{-t}dt$.

For a fading channel, the effects of channel must be considered, and Willie estimates the existence of communication using a threshold determined by channel state information (CSI), which is assumed to be perfectly known in advance.
In this case,
the optimal threshold corresponds to the value minimizing $P_{e,w}\qty(\lambda)$ and is expressed as~\cite{Shah2021fading}
\begin{align}
    \lambda_{\text{CSI}}= \frac{\gamma^{-1}_{w}\left(\abs{h_{w}}^{2}+ \gamma^{-1}_{w}\right)}{\abs{h_{w}}^{2}}\log\left(\frac{\abs{h_{w}}^{2}+ \gamma^{-1}_{w}}{\gamma^{-1}_{w}}\right). \label{eq:lmd_CSI}
\end{align}

\section{Benchmark Coherent Schemes}
\label{sec:conv}
In this section, we introduce coherent Gaussian signaling schemes as benchmarks
for our proposed system.
Then, we analyze the effect of channel estimation errors on the achievable rate
when using Gaussian-distributed pilot symbols.

\subsection{Coherent Gaussian Signaling Schemes}
Gaussian signaling~(GS) is a modulation scheme that transmits data symbols following
a Gaussian distribution. The first scheme, PSK-based GS (PSK-GS), is
essentially equivalent to the one proposed in \cite{katsuki2023NGS}, and the
second scheme, chaos-based GS (chaos-GS) \cite{2012OkamotoChaos}, is superior
in terms of performance when the transmission rate is equivalent to BPSK.

\subsubsection{PSK-GS \cite{katsuki2023NGS}}
This scheme is a special case of the conventional differential scheme of
\cite{katsuki2023NGS}, which relies on the property that the probability distribution
does not change when a data symbol distributed on a unit circle is multiplied
by a random variable following a specific distribution due to its circular symmetry.
Let $M$ denote the modulation order of PSK symbols and let
$\mathcal{X}=\qty{e^{j\frac{m}{M}\pi}\mid m=0,\cdots,M-1}$ denote the
constellation set. The $k$th modulated symbol is generated as $s_{k}=g \cdot e^{j\frac{m_k}{M}\pi}$, where $g$ is the Gaussian RV based on a pre-shared seed.
The Gaussian RV $g$ can also be varied as $g_k$ depending on the index $k$.

The fundamental limitation of PSK-GS is the need for Alice and Bob to have
exactly the same seed value, which must be perfectly shared in advance. This pre-shared
seed enables modulation and demodulation. If the seed value is even slightly
incorrect, the performance will be significantly degraded.

\subsubsection{Chaos-GS \cite{2012OkamotoChaos}}
This scheme is based on chaos theory and efficiently generates symbols that
follow a Gaussian distribution. Originally proposed as a secure modulation scheme,
it is readily applicable to covert communication scenarios.
As with PSK-GS, this scheme
also requires the pre-sharing of a seed value, which enables coherent detection of
the received symbols.
The detailed construction process is also described in
\cite{ishikawa2021artificially}.\footnote{The open-source implementation of
    chaos-GS is available at \url{https://github.com/ishikawalab/wiphy/blob/master/wiphy/examples/okamoto2012chaos.py}}

In both cases,
the transmit symbol vector is estimated by the maximum likelihood detector~(MLD) $\hat{\s} = {\arg\min}_{\s \in \mathcal{S}} \| \y_{b}-\s \hat{h}_{b}\|^{2}$, where $\mathcal{S}$ denotes a GS codebook constructed with a seed shared without error for generating RVs.

\subsection{Channel Estimation with Gaussian-Distributed Pilots}

In typical communication systems, the transmission power allocated for pilot symbols is higher than
the one for data symbols to enable accurate channel estimation.
By contrast, in the context of covert communication with channel estimation,
the DEP can be minimized when the transmission power is equally distributed between pilot and data symbols~\cite{Xu2019PilotBasedCE}, which is the assumption followed in this letter as well.

The estimated channel considering estimation errors is $\hat{h}=\sqrt{1-\beta^{2}}h+\beta e$~\cite{endo2024Grassmann}, where $e\sim\CN\qty(0,1)$ is an error 
and $\beta\in\qty[0, 1]$ is the uncertainty of estimation.
The estimation error can be evaluated by the normalized minimum square error~(NMSE) given by
\begin{align}
    \sigma^{2}_{e} \triangleq \E\qty[\abs{\hat{h}-h}^{2}] ~ / ~ \E\qty[\abs{h}^{2}],
    \label{eq:NMSE_def}
\end{align}
and that calculated by MMSE equalization with known pilot symbols is expressed as $\sigma^{2}_{e} = 2\cdot (1-\sqrt{1-\beta^{2}}) = (1+\gamma_{b}\norm{\p}^{2})^{-1}$~\cite{endo2024Grassmann}.

We use Gaussian-distributed pilot symbols in this letter,
as they can maintain the achievable rate of
the coherent GS schemes while ensuring maximum covertness.
Here, we assume that the generated pilot symbols vary with each instance $\mathcal{H}_{1}$ and behave as i.i.d Gaussian RVs. NMSE is calculated by the numerical integration of
\begin{align}
    \sigma_e^2 = \E_{\p} \qty[\qty(1+\gamma_{b}\norm{\p}^{2})^{-1}]
    \label{eq:NMSE-expectation}
\end{align}
over the Erlang distribution
\begin{align}
    f\qty(\norm{\p}^{2}) = \frac{\norm{\p}^{T_{P}-1}}{\qty(T_{P}-1)!}\exp\qty(-\norm{\p}^{2}), \label{eq:PDF_p}
\end{align}
which has not been investigated in existing studies~\cite{katsuki2023NGS,Shah2021fading}, and is newly developed for evaluating the channel estimation error in the benchmark coherent schemes.

\begin{figure}[t]
    \centering
    \includegraphics[width=0.95\linewidth, keepaspectratio]{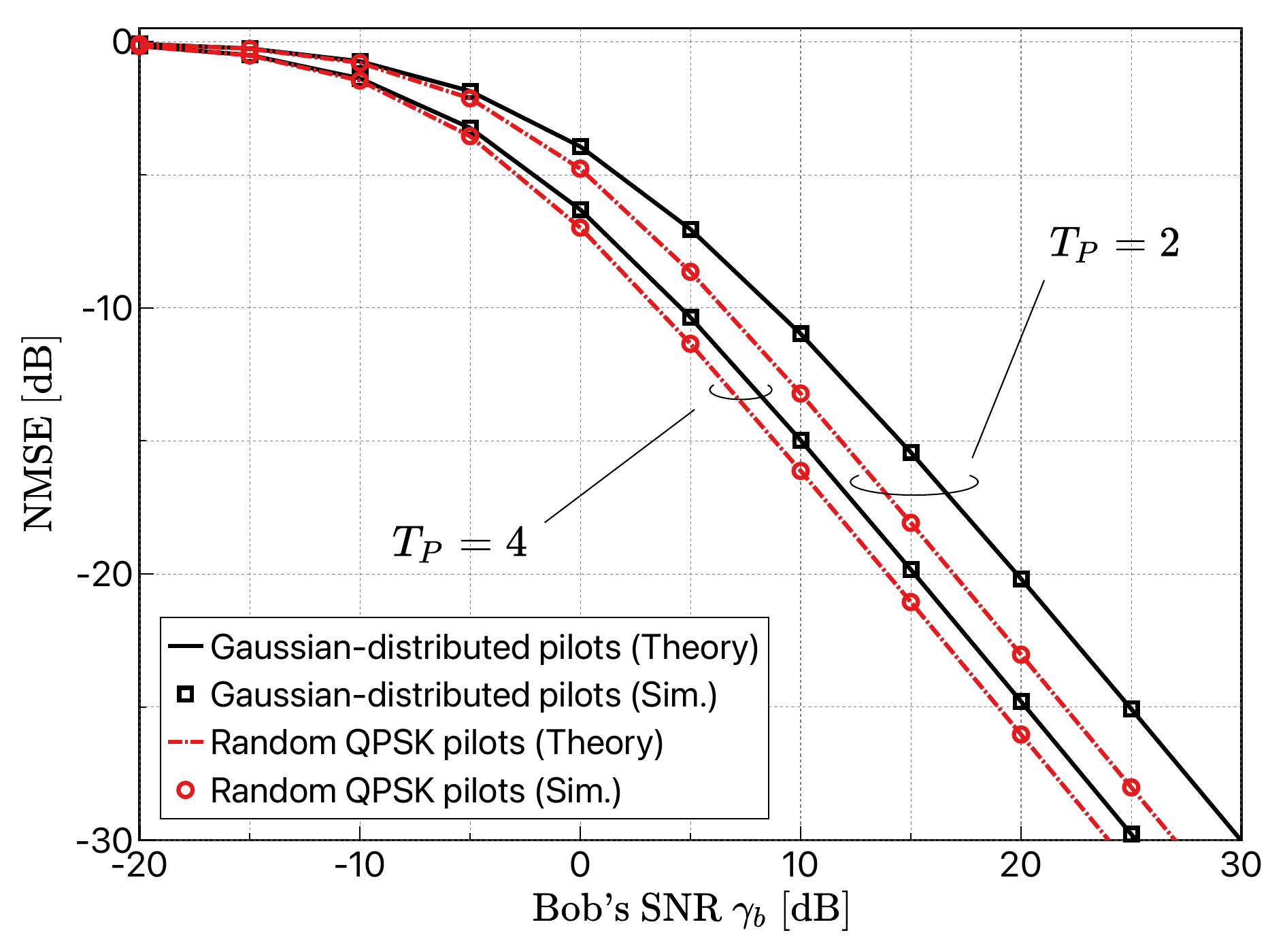}
    \vspace{-1ex}
    \caption{Comparisons of channel estimation accuracy.}
    \vspace{-2ex}
    \label{fig:NMSE_GRS}
\end{figure}
As an example, Fig.~\ref{fig:NMSE_GRS} shows a comparison of channel estimation errors induced by MMSE with Gaussian-distributed pilot symbols as well as the random QPSK symbols defined in 5G NR, where the simulation \eqref{eq:NMSE_def} and theoretical values \eqref{eq:NMSE-expectation} were considered.
As shown in Fig.~\ref{fig:NMSE_GRS},
the simulation and theoretical values matched perfectly, indicating that the theoretical analysis is correct.
Also, the channel estimation accuracy improved as the number of Gaussian pilot symbols increased, although it was inferior to the reference QPSK case.

\subsection{Achievable Rate Assuming Channel Estimation Errors}
The ideal transmission rate per slot when transmitting $B$ bits over $T$ slots can be defined as $R \triangleq B/T$, and the achievable rate corresponding to a benchmark coherent scheme is given by \cite{endo2024Grassmann}
\begin{align}
    \!R_{e}
    \!=\!
    R\!-\!\frac{1}{T\mathcal{\abs{S}}}\mathbb{E}_{\hat{h},\n}\!\qty[\sum^{ \abs{\mathcal{S}}}_{i=1}\log_{2}\qty(\!\frac{\sum^{\abs{\mathcal{S}}}_{j=1}p\qty(\y_{i}\mid\hat{h},\s_{j})}{p\qty(\y_{i}\mid \hat{h},\s_{i})})\!] ,
    \label{eq:achievable_rate_benchmark}
\end{align}
where $\y_{i}=\s_{i}\hat{h}+\n$ is a received symbol vector given a codeword $\s_{i}\in \mathcal{S}$ and the likelihood ratio is expressed as $p(\y_{i}\mid \hat{h},\s_{j}) / p(\y_{i}\mid \hat{h},\s_{i}) =
\mathrm{exp} [(-\norm{\n_{i,j}}^{2} + \norm{\n_{i,i}}^{2}) / (\sigma^{2}_{b}+\sigma^{2}_{e})]$ with $\n_{i,j} = \y_{i}-\s_{j}\hat{h}$.

\section{Proposed Noncoherent Covert Communication}
\label{sec:prop}
In this section,
we design GS using Grassmann codewords,
which is referred to as \textit{Grassmann-based GS~(Grass-GS)},
for noncoherent covert communication.
Then, we analyze the achievable rate when using the proposed Grass-GS.

\subsection{Design of Grass-GS}
Following the coherent system model of \eqref{eq:yb}, the transmit symbol vector is denoted by $\s = \qty[s_{1} ~ \cdots ~ s_{T}]$, since Bob can decode information without channel estimation.
That is, we set $T_{P}=0$ and $T_{D}=T$ here to provide fair comparisons between both coherent and noncoherent schemes, as illustrated in Fig.~\ref{fig:frame}.
A well-known method constructs a Grassmann constellation using the matrix exponential map \cite{kammoun2007exp}. Another method constructs it to be uniformly distributed on the Grassmann manifold \cite{ngo2020cubesplit} based on numerical optimization~\cite{Nicolas2014manopt}, termed \textit{original manopt constellation}.

In constructing the manopt constellation, if the starting point for optimization is set randomly and a simple gradient descent method is used, the phases of the optimized constellation generally follow a uniform distribution.
Here, by imposing a constraint in the optimization process that all symbols have constant amplitude,
it is possible to generate symbols that follow an arbitrary distribution in a manner similar to PSK-GS,
at the expense of a reduced minimum distance of codewords.

Let $\mathcal{G}\qty(T,1)$ denote the Grassmann manifold, representing the set of all 1-dimensional subspaces in $\mathbb{C}^T$, and let $\c_{i} \in \mathcal{C} \in \mathcal{G}\qty(T,1)$ represent a Grassmann codeword chosen from the codebook $\mathcal{C}$ of size $2^{B}$.
In the design of Grass-GS, we use the minimum chordal distance~(MCD), 
because it is a standard design metric and can be approximated by the log-sum-exp function.
But, essentially, any design metric can be used.
The design of the constant-amplitude Grassmann constellation can then be formulated as
\begin{equation}
    \begin{aligned}
        \minimize_{\mathcal{C}}\quad & \mathrm{log}\sum_{1\leq i < j \leq |\mathcal{C}|}\exp\left( \frac{|\mathbf{c}^{\mathrm{H}}_i\mathbf{c}_j|}{\epsilon}\right) \\ \text{s.t.}\quad&\mathbf{c}_{i}\in \mathcal{G}(T, 1),\;\forall\; i=\{1,\cdots,|\mathcal{C}|\}, \\&|c_{i,t}| = 1, \;\forall\; t=\{1,\cdots,T\},
    \end{aligned}
    \label{eq:opti_lse_M1}
\end{equation}
where $\epsilon$ represents a smoothing constant and is set to $\epsilon=10^{-2}$ in this letter.

The proposed Grass-GS codeword is generated by multiplying the codeword $\c_i \in \mathcal{C}$ by a Gaussian RV at Alice, i.e.,
\begin{align}
    \s = g \cdot \c_{i}.
    \label{eq:s}
\end{align}

At Bob, MLD is performed on the received symbols.
Since~$g$ is unshared with Bob, 
its product with the channel coefficient~$h_{b}$ yields the equivalent channel which includes the effect of $g$,
and the likelihood function of the received symbol given $\c_{i}$ is expressed as
\begin{align}
    p\qty(\y_b \mid \c_{i})
     & 
     =\frac{\exp \qty(-\operatorname{tr}\qty{\y_b^{\mathrm{H}}\qty(\mathbf{I}_T+\c_{i}\c_{i}^{\mathrm{H}})^{-1} \y_b})}{\pi^{T} \operatorname{det}\qty(\mathbf{I}_T+\c_{i} \c_{i}^{\mathrm{H}})} \\
     & =
    \frac{\exp \qty(-\norm{\y_{b}}^{2}+\frac{1}{1+\norm{\c_{i}}^{2}}\abs{\y_{b}^{\mathrm{H}} \c_{i}}^{2})}{\pi^T\qty(1+\norm{\c_{i}}^2)}.
\end{align}
Thus, the received symbol is estimated as follows:
\begin{align}
    \hat{\c} =
    \underset{\c \in \mathcal{C}}{\arg\max}\abs{\y_{b}^{\mathrm{H}} \c}^{2},
    \label{eq:GLRT}
\end{align}
where the RV $g$ multiplied at Alice is no longer required.
The benchmark schemes, PSK-GS and Chaos-GS,
require the legitimate transceiver to share the random seed, while the proposed Grass-GS can avoid the sharing process that diminishes covertness.
Note that all the considered schemes still require the pre-sharing of the codebook.

\begin{figure}[t]
    \centering
    \includegraphics[width=0.95\linewidth,keepaspectratio]{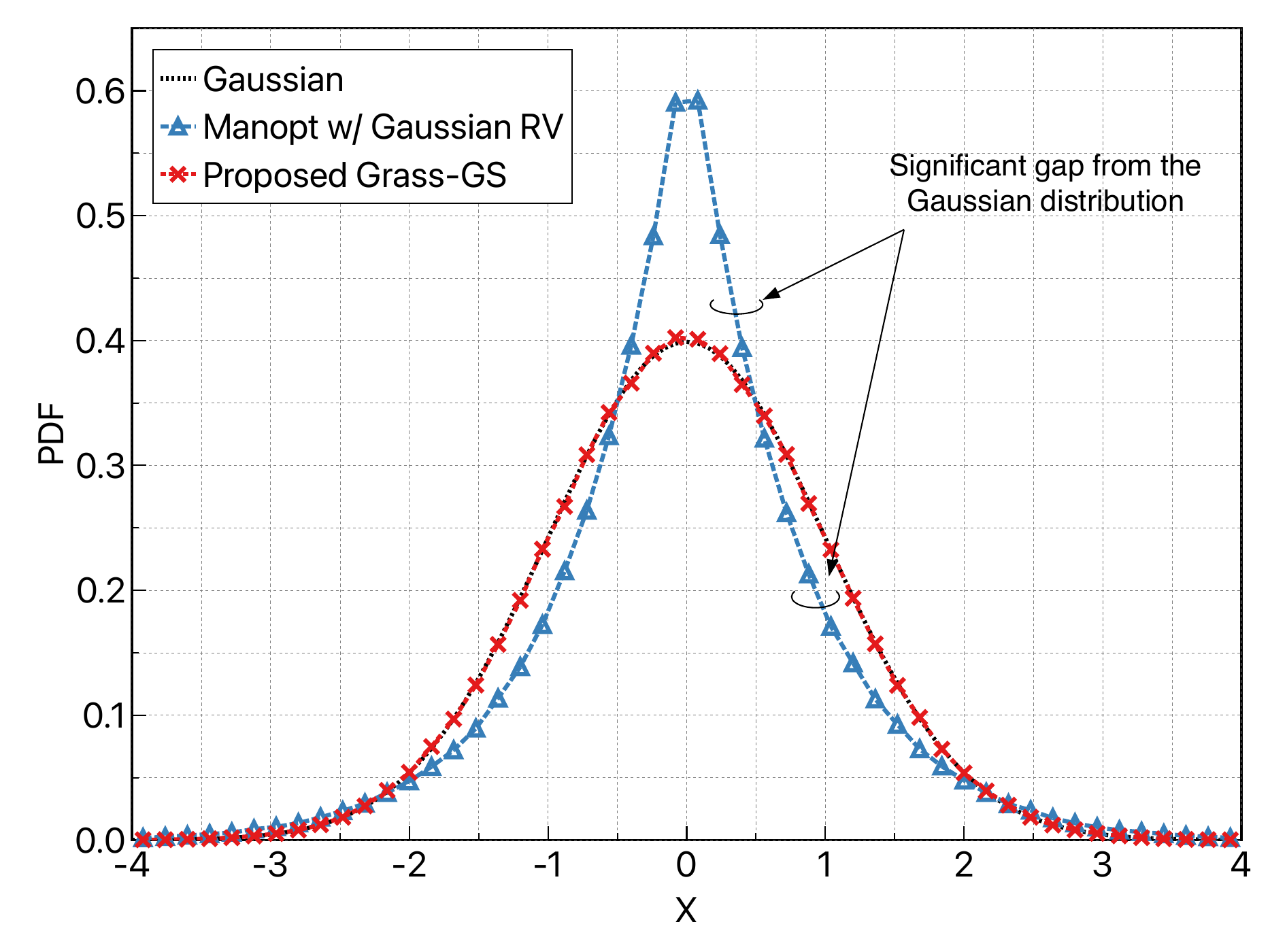}
    \vspace{-1ex}
    \caption{PDF of the real or imaginary component of generated symbols.}
    \vspace{-2ex}
    \label{fig:gaussianity}
\end{figure}
To evaluate Gaussianity,
Fig.~\ref{fig:gaussianity} shows the probability distribution function (PDF) of the real or imaginary component of the proposed Grass-GS, where the parameters were set to $T=4$ and $B=2$.
Additionally, we also considered a naive noncoherent GS schemes which multiplies the original manopt constellation
by a Gaussian RV.
As shown in~Fig.~\ref{fig:gaussianity}, the proposed Grass-GS exactly followed a Gaussian distribution, which was self-evident.
Also, the naive noncoherent GS schemes, which did not impose the circular symmetry constraint, deviated significantly from the Gaussian distribution.

\subsection{Detection Error Probability at Willie}
Unlike the coherent scenario in Section~\ref{subsec:dep} with the assumption of perfect CSI at Willie, in the noncoherent scenario,
the existence of communication is estimated at Willie using channel distribution information (CDI) since he cannot access CSI due to the absence of pilots.
In this case, the optimal threshold is obtained by employing the numerical search of
\begin{align}
    \lambda_{\text{CDI}}
    =\underset{\lambda^{*}}{\arg\min}~\mathbb{E}_{\abs{h_{w}}^{2}}\qty[P_{e,w}\qty(\lambda^{*})],
    \label{eq:lmd_CDI}
\end{align}
where the expectation is taken over the PDF of $\abs{h_{w}}^{2}$.

\subsection{Achievable Rate of Grass-GS}
The achievable rate of Grass-GS is given by~\cite{endo2024Grassmann}
\begin{align}
    R_{g}
    \!=R-\!\frac{1}{T\abs{\mathcal{C}}}\mathbb{E}_{h,g,\n}\qty[\sum^{\abs{\mathcal{C}}}_{i=1}\log_{2}\qty(\frac{\sum^{\abs{\mathcal{C}}}_{j=1}p\qty(\y_{i}\mid\c_{j})}{p\qty(\y_{i}\mid\c_{i})})],
    \label{eq:achievable_rate_proposed}
\end{align}
where $\y_{i}=\c_{i} \cdot g \cdot h + \n$ is the received symbol when $\c_{i}\in\mathcal{C}$ is transmitted and the likelihood ratio is
\begin{align}
    \frac{p\qty(\y_{i}\mid \c_{j})}{p\qty(\y_{i}\mid \c_{i})}
    =\exp\qty(\frac{\abs{\y^{\mathrm{H}}_{i}\c_{j}}^{2}-\abs{\y^{\mathrm{H}}_{i}\c_{i}}^{2}}{\sigma^{2}_{b}\qty(T+\sigma^{2}_{b})}).
\end{align}
Since we have $\mathbf{y}_i = \mathbf{c}_i \cdot g \cdot h + \mathbf{n}$, this expression is not strictly identical to the one given in \cite{endo2024Grassmann}.
Due to the constant-amplitude constraint, MCD decreases compared to the original manopt constellation, which is a disadvantage of the proposed approach.

\section{Performance Comparisons}
\label{sec:comp}
In this section,
we evaluate the performance of the benchmark coherent schemes, PSK-GS \cite{katsuki2023NGS} and Chaos-GS \cite{2012OkamotoChaos}, and the proposed noncoherent scheme, Grass-GS, in terms of the DEP at Willie and the achievable rate at Bob.

\subsection{Detection Error Probability}
\label{subsec:covertness}
\begin{figure}[t]
    \centering
    \includegraphics[width=0.95\linewidth, keepaspectratio]{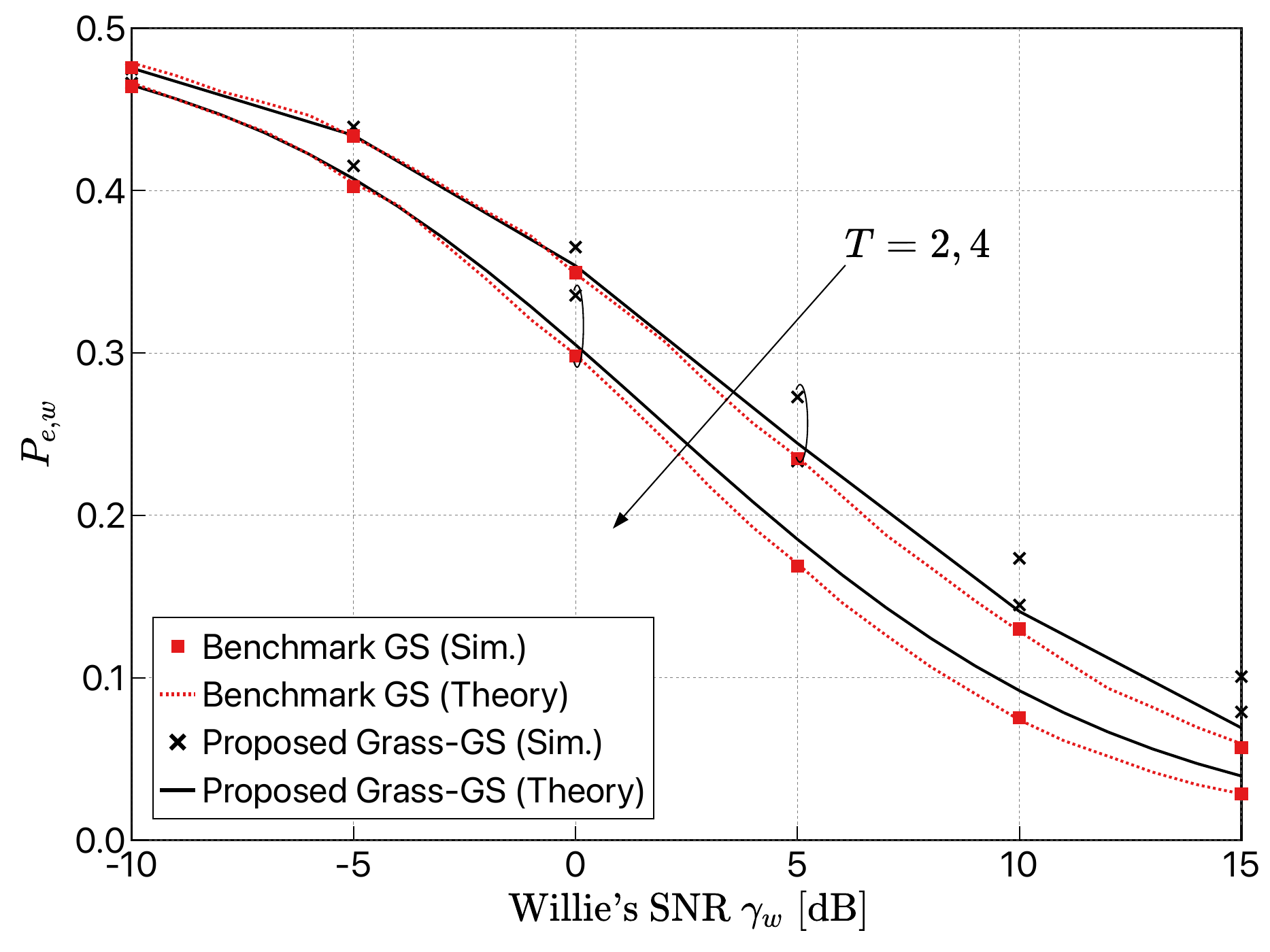}
    \vspace{-1ex}
    \caption{Comparison of DEP at Willie that corresponds to the covertness.}
    \vspace{-2ex}
    \label{fig:P_eW}
\end{figure}
First, Fig.~\ref{fig:P_eW} shows DEP comparison upon increasing the received SNR at Willie,
where the detections of the benchmark coherent and the proposed noncoherent GS were performed using the thresholds evaluated with \eqref{eq:lmd_CSI} and \eqref{eq:lmd_CDI}, respectively.
The number of time slots was $T=2$ or $4$ in both cases.
For the benchmark coherent GS,
pilot and data symbols were used for detection.
As shown in Fig.~\ref{fig:P_eW},
the proposed scheme outperformed the benchmarks in all SNR region,
but it surpassed the theoretical curve as the SNR increased.
This gap occurred due to the time correlation in the received symbols,
implying that the proposed GS might be detected by evaluating the correlation.
By contrast, the gap narrowed as the length $T$ was decreased. Then, short data symbols are favorable for maintaining robustness against a correlation-based detection.
Overall, the proposed noncoherent GS remains effective for the radio meter detection,
particularly when compared with the coherent benchmarks.

\subsection{Achievable Rate Comparisons}
\label{subsec:AMI}
\begin{figure}[t]
    \centering
    \includegraphics[width=0.95\linewidth, keepaspectratio]{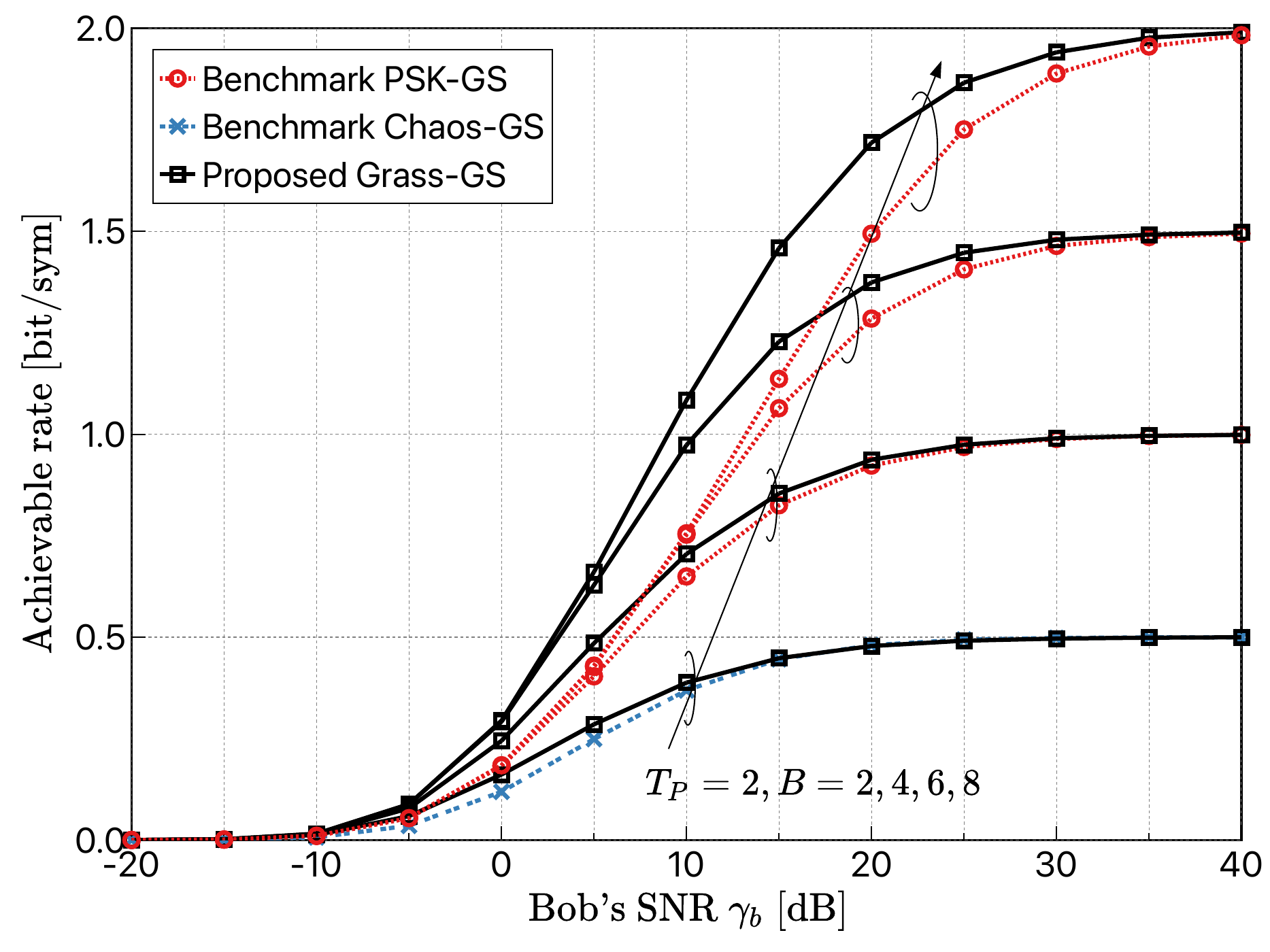}
    \vspace{-1ex}
    \caption{Achievable rates of the benchmark and proposed GS schemes.}
    \vspace{-2ex}
    \label{fig:AMI}
\end{figure}
Next, Fig.~\ref{fig:AMI} compares the achievable rates of the benchmark and the proposed GS.
The lengths of data and pilot symbols were $T_{D}=T_{P}=2$, resulting in the total length of $T=4$.
The number of the transmission bits was set to $B=4$, $6$ or $8$ for PSK-GS and $B=2$ for Chaos-GS.
The channel estimation error in the benchmark schemes
was evaluated with \eqref{eq:NMSE-expectation}.
Note that both schemes were fairly compared in terms of $R$.
As shown in Fig.~\ref{fig:AMI}, the proposed Grass-GS exhibited similar performance to Chaos-GS at low rates and significantly outperformed PSK-GS at high rates, achieving its half-rate at an SNR up to 4.7 dB lower than the coherent benchmarks. This means that Grass-GS can accordingly reduce the transmit signal power, thereby offering a substantial advantage in enhancing covertness. Note that similar trends were observed for a smaller pilot ratio $T_{P}/T=4/14 \approx 0.286$. Although bit error rate comparisons were omitted, Grass-GS achieved higher diversity orders than the coherent benchmarks.

\subsection{Complexity Analysis}

\begin{table}[t]
    \centering
    \caption{Complexity comparisons.}
    \vspace{-1ex}
    \begin{tabular}{|l||l|l|}\hline
                                               & Number of RVs        & Decoding complexity                          \\
        \hline
        Conv. PSK-GS \cite{katsuki2023NGS}     & $\mathcal{O}(T)$     & $4 \cdot 2^{B/T} T = \mathcal{O}(2^{B/T} T)$ \\
        Conv. Chaos-GS \cite{2012OkamotoChaos} & $\mathcal{O}(2^B T)$ & $4 \cdot 2^{B} T = \mathcal{O}(2^{B} T)$     \\
        Prop. Grass-GS                         & $\mathcal{O}(1)$     & $4 \cdot 2^{B} T = \mathcal{O}(2^{B} T)$     \\
        \hline
    \end{tabular}
    \vspace{-2ex}
    \label{tab:comp}
\end{table}
Finally, we compare the computational complexity required for the three considered schemes.
Specifically, for $T$ time slots, we derived the complexities for generating RVs and performing MLD, summarized in Table~\ref{tab:comp}.
Evidently, the conventional coherent benchmarks require additional complexity for channel estimation, but since there are a number of methods available, it was omitted here.
As given, the proposed Grass-GS requires the lowest number of RVs while maintaining the same decoding complexity as the benchmarks, making covert communications more practical.

\section{Conclusions}
\label{sec:conc}
We have proposed a noncoherent covert communication scheme that employs constant-amplitude Grassmann constellation to generate GS.
For fair comparisons,
the benchmark coherent schemes have been introduced,
and the achievable rate under channel estimation errors when using Gaussian-distributed pilot symbols has been evaluated.
Our performance comparison demonstrated that the proposed Grass-GS achieved both higher covertness and achievable rates than the coherent benchmarks.
Future work will focus on a MIMO system and its optimal detection at Willie.

\footnotesize{ \bibliographystyle{IEEEtranURLandMonthDiactivated} \bibliography{main} }
\end{document}